# A Model Driven Approach on Object Oriented PLC Programming for Manufacturing Systems with regard to Usability

Martin Obermeier, Steven Braun, and Birgit Vogel-Heuser, *Senior Member, IEEE*

*Abstract*—This paper presents the modular automation for reuse in manufacturing systems (modAT4rMS) approach to support the model-driven engineering (MDE) of object oriented manufacturing automation software with regard to its usability and software modularity. With usability we refer to the aspects effectiveness, efficiency and user acceptance, as defined by ISO 9241-11. The modAT4rMS notations are based on selected features from the Unified Modeling Language (UML) and the Systems Modeling language (SysML) and iteratively further developed by a series of empirical studies with industrial practitioners as well as mechatronics trainees. With modAT4rMS a MDE approach for Programmable Logic Controller (PLC) programming was developed with the goal to facilitate modular object oriented programming of PLC software by improving the representation of the relationships between the structure and behavior diagram types and by reducing the level of abstraction in the structure model. modAT4rMS notations for PLC software structure and software behavior modeling are presented and illustrated with a modeling example using a modAT4rMS editor prototype.

For the evaluation of the developed notations the results from a study with 168 participants is presented, showing the benefits of this new approach in comparison to the classic procedural paradigm (IEC 61131-3) and the domain specific UML profile plcML in regard to programming performance and usability aspects. Finally the advantages and limitations of the approach are discussed and an outlook for further development is given.

## I. INTRODUCTION

In machine and plant automation model-driven object-oriented (OO) software engineering of PLCs remains a challenge [1]. Since the IEC 61131-3 standard [2] is widely accepted in industrial automation [3], using OO features for software model creation in this field is still uncommon [4], [5]. Reasons seem to be manifold: absence of accepted modern PLC software engineering approaches, classical procedural programming languages on the PLCs [2], mostly application engineers coming from application domains and also the fact that the integration of OO features into the IEC61131-3 in 2013 is still relatively new and still only supported by few PLC developer environments, like 3S Codesys V3 [5] and Beckhoff TwinCAT 3 [6].

As Vyatkin mentions in [5], the adaption of developments from the general computing allow the developers to take advantage of the huge investments into such technologies and rely on proven solutions, like OO, MDE or component-based design, rather than reinventing the wheel.

In order to increase reuse and quality in manufacturing system (MS) software engineering, modeling techniques from computer science can be applied in automation and control engineering, e.g. Unified Modeling Language (UML) and Systems Modeling Language (SysML). In computer science, object oriented programming has been considered a solution to many computing challenges. As Mili et al. [7] describe, "Software engineers view object orientation (OO) as the answer to their numerous and intractable problems: enhancing software quality, reusability, and providing a seamless development methodology". In order to support software engineers with regard to usability aspects, UML has also already been successfully applied, cf. [8]. It is most likely that application and process control engineers will program their solutions using such techniques in the future, as e.g. Thramboulidis and Frey [3] consider the benefit of integrating IEC 61131-3 with UML being able to use several diagrams in order to capture more aspects of the system. This is supposed to result in a more complex yet still comprehensible model.

This article presents the newly developed modular automation for reuse in manufacturing systems (modAT4rMS) approach which adapts strengths and resolves the weaknesses of notations in PLC software engineering for plant automation, e.g. UML, SysML and the classical IEC 61131-3 languages. The focus of this work is classical manufacturing automation, i.e. open loop control, mostly implementing interlocking algorithm, feedback, PID and other, simple closed loop controllers. Three studies with more than 100 participants (including industry practitioners, mechatronic trainees and researches) have been conducted in the course of its development. For its evaluation a study with 168 participants (mechatronic trainees and industrial practitioners) was carried out, in order to test modAt4rMS with regard to usability against the state of the art IEC 61131-3 and an UML-based

MDE approach (plcML). The reason for the evaluation against IEC 61131-3 FBD is evidently the need for a well-known comparison basis from industrial practice for empirical experimentation. The plcML, as described in II.A.2 and III, is a predecessor of the modAt4rMS and was therefore used as a comparison approach. In the following chapter the state of the art on modeling notations and a short overview of the previously conducted studies are presented. The challenges when adapting UML to design modular machine software are shown and necessary steps in order to facilitate the use of UML or domain specific UML profiles for modeling the structure and behavior of automation systems are derived, resulting in the modAT4rMS notations. In chapter III the concepts of the modAT4rMS behavior and structure notations are shown. The latest study, evaluating the modAT4rMS is discussed in short in Chapter IV and its results are presented. Finally, the article is concluded in chapter V and an outlook on future work is given.

## II. Related Work

This chapter covers relevant notations in the context of PLC programming followed by related results from previously conducted studies.

### A. Notations for PLC programming

In the following, related aspects on the widely used IEC 61131-3 [2] standard languages will be discussed first and in contrast thereof the UML and its relevant domain specific profiles will be covered.

*1) IEC 61131-3*

The IEC 61131-3 [2] with its textual (Structured Text (ST), Instruction List (IL)) and its graphical languages (Ladder Diagram (LD), Function Block Diagram (FBD)) as well as its Sequential Function Chart (SFC) is still accepted and widely used by software engineers to specify the software part of industrial automation systems [9].

The clean encapsulation of code and data in function blocks is possible, but is often disturbed by the use of global variables, which leads to poor reusability of modules, cf. [10]. In order to provide an elemental basis of reuse, IEC 61131-3 software can be structured into Program Organization Units (POUs), e.g. function blocks (FBs). FBs capture the structure and behavior of a collection of objects used in automation projects [9] and can be regarded as a specific type of class being instantiated before the system's operation. Hajarnavis and Young [11] evaluated an object oriented approach in comparison to IEC 61131-3 approaches in an experiment with 63 participants. Their results showed, that "there is a difference in completion rate and time between the programming tools under test. Although the object-oriented programming tool used here proved best suited to the facilitation of fast and correct process changes, ladder logic based on a step-based structure was found to give comparable results when used by skilled programmers.", cf. [11].

As Fantuzzi proposes in [12], a modular quasi object oriented program structure is possible, through the application of design patterns. In [5], Vyatkin mentions the centralized programmable control model of PLCs as a critical aspect as it could "cause unpredictable results and overheads when two or more controllers communicate via network, and its centralized, cyclically scanned program execution model limits the reuse of the components in case of reconfiguration."

*2) UML and domain specific UML profiles for MS*

UML is an established standard for software development being used to create models from a conceptual basis or software perspective. The current UML specification (V2.4.1 [13]) contains 16 diagrams providing both a structural view, e.g. class diagram, defining the dependencies between FBs or classes, and a behavior view, e.g. state machines and activity diagrams, easing the implementation of behavioral aspects of the system. The object-oriented paradigm encourages a modular design by different relation types, such as association among classes and inheritance for hierarchical structuring. The UML class diagram illustrates the structure of automation control software and helps to understand the interdependencies of components.

Thramboulidis and Frey [9], Thramboulidis [14], Secchi et al. [15], [16] and Basile et al. [17] highlighted the benefit of using modeling notations through improving modularity and reuse as well as system understanding compared to traditional approaches using pure IEC 61131-3 programming languages. In [18] alongside a model checking approach using UML, a UML-based hardware/software co-design platform for embedded systems using FPGA devices is proposed in order to support the direct interaction between the UML models and the real hardware architecture.

However, Secchi et al. [16] mentioned that it is not sufficient to employ UML with its 16 diagrams but to provide an adapted modeling language for domain-specific applications, i.e. UML-RT, in order to provide modeling of real-time software architectures by means of encapsulated software parts interacting with each other through ports following a communication protocol.

Kim et al. present in [19] an state chart approach for embedded systems that includes the necessary time and resources for actions in the state chart model in order to describe the competition between processes using shared resources.

In [20] Bicchierai et al. present an MDE approach adapting UML for Modeling and Analysis of Real-Time and Embedded systems [21], with a selected number of diagrams (component, class, object and activity diagrams), for managing a documentation process and in combination for providing guidance for its translation into preemptive Time Petri Nets.

In order to model a holistic view of the automation system considering all disciplines involved in engineering, the SysML may be used [9], [22]–[24]. Hästbacka et al. [25] use the UML Automation Profile in their AUKOTON approach in order to provide domain specific modeling constructs for a holistic view including requirements, automation components and devices as well as the process system and the control systems. As this article focuses on the PLC software part of the system represented in notations, related work specific to this aspect is discussed further.

In several works [23], [26]–[30] the attempt to integrate IEC 61131-3 notations with UML or SysML have been made. Yet, previous research on software engineering with UML shows that while using more diagrams is confusing [23], using adapted subsets of UML is beneficial, cf. [27].

Therefore and based on the findings in [27] the plcML, a domain-specifically adapted UML profile for PLCs, was developed by Witsch, cf. [31], [32]. The plcML reduces the number of notations to three, offering the class diagram for structure modeling and adapted activity and state chart diagrams for behavior modeling. These diagrams are described in detail in [31], [32] where also their model transformation using OCL is described as well as their model formalization, in order to enable code generation and bidirectional mapping for IEC 61131-3 development environments. Editors for the plcML notations and the corresponding code generation are in the making as a plugin for CODESYS V3 and will provide a fully real-time capable plcML programming environment (product name CODESYS Prof. Developer Edition - CODESYS UML), cf. [32].

The presented related work on IEC 61131-3 and UML approaches lead to the development of the new modAT4rMS approach with the following basic conclusions on the notation's design: Only a limited number of domain specific diagrams should be provided, not the full UML/SysML. The notation should support effective and efficient OO MDE with regard to user satisfaction in comparison to the state of the art IEC61131-3 approach, as this is a major focus of MDE, cf. [5]. These conclusions were refined through the results of a series of previously conducted studies, described in the following section II.B.

*B. Results from previously conducted studies*

In a series of previously conducted studies the influence of different factors of notations on the usability for PLC programming were evaluated and the new modAT4rMS notations for structure and behavior modeling were developed. For these studies the usability was measured according to the aspects effectiveness, efficiency, user acceptance defined in ISO 9241-11 [33]. For the measurement of the effectiveness, the task completion was determined using code review, supported by review guidelines which included all programming tasks with their respective subtasks down to the creation of an attribute/variable of the structure model respectively a function/operation call in the behavior model. Each review on task completion was carried out by two evaluators in order to verify the objectivity of the evaluation method. The level of task completion was then calculated as the percentage of solved tasks over all given tasks.

In a first study, conducted in 2011 and presented in [4], [34], the model driven plcML approach's state chart and class diagrams, cf. [31], [32], were experimentally evaluated in comparison to the IEC 61131-3 FBD. The results of this study showed no significant differences in programming performance between FBD and plcML. The required level of abstraction for creating classes proved problematic for the participants, who needed tendencially more time for structure creation, (time measured until they started creating the behavior model: $m_{\text{IEC 61131-3}} = 49.83$ minutes; $m_{\text{plcML}} = 56.05$ minutes, probability of results occurring by chance $p = 0.06$) and created superfluous classes, cf. [4]. From these results a need for an adapted, less abstract structure notation, supporting the creation of modular software was deducted.

Therefore, a first draft of the modAT4rMS notations was developed, which used an object centered structure diagram and tried to visualize the connection between structure and behavior diagram by using an adapted UML state chart for behavior modeling, cf. [35]. This draft was evaluated in a 4 hours study with 10 researchers of the Institute for Automation and Information Systems, at the Technische Universität München. In this study the participants first had to program a known automation lab demonstrator, using only the new draft notations. In the second part a given model had to be interpreted. Overall 35 minutes of training were provided, 45 minutes for solving the model creation task (structure and behavior) and 35 minutes for the interpretation task were given. After solving all test tasks, the participants had to answer questionnaires with scaled and open answer possibilities on the subjective usability, adapted from [33], [36], and took part in a focus group discussion on the new notations. The test task results showed a mean task completion level of 68.5% for structure and 59.5% for behavior model creation. The mean result on the interpretation task was 77.66% completion. Further results from the questionnaires and the focus group were positive for the developed structure notation and showed that the behavior notation in this draft was rated better than UML state chart diagrams. Yet the behavior diagram was criticized for being unclear in regard to the connection between structure and behavior for seasoned UML modeling experts. The structure model had worked very well with a few exceptions and was therefore developed further on this base. For clearer connections between the blocks of the structure and the processes that shaped the software model, the behavior notation was changed in the next draft, which subsequently led to the final modAT4rMS notations.

In the second draft the behavior notation was changed to a notation adapting plcML activity diagram and UML sequence diagram features. This proved to solve the criticized lack of clarity of the connection between structure and behavior diagram in the second evaluation, being conducted with eight industrial PLC programming experts and nine trainees on electronics. In this study the participants had to program a known automation system used for the apprentices' PLC programming training, with the second draft modAt4rMS notations. Like in the previous study the same questionnaires were used for measurement of the subjective usability and workload. The results of this study led to the final version of the modAT4rMS notations described in III.

### III. THE MODAT4RMS APPROACH

From the related work on modeling notations it can be deduced that the strength of UML and SysML to represent many different aspects of the development of a mechatronic system, is not optimally applicable for the user in the specific case of PLC software engineering for MS. This is reflected in the discussed previous evaluation results in II, which pointed out especially the criticized high complexity of the class diagram and the problem of representing relationships between the structure and behavior diagram types as the key aspects for usability improvement.

With the modAT4rMS approach a solution for these aspects for central PLC systems is proposed. In general, the modAT4rMS approach is intended to guide the user from the first step on through the modeling process to gain a highly modularized MS structure, including the functional relations of the system. Among other aspects this is targeting to improve the detection of similar or identical objects, which proved difficult in previous studies, cf. II.B. Further central aspects are to reduce the required level of abstraction skills by the use of less abstract blocks instead of abstract classes and by the proposition of a strictly limited level of inheritance. In this section the modAT4rMS structure and the behavior models are presented. As formal definitions are not the focus of this paper and formalizations and transformations of similar diagrams have already been presented in several publications [18], [19], [26], [28], [29], [32], [37], only relevant meta-model descriptions and their connections to the UML and plcML are given for each notation showing the changes made in order to improve the effectiveness, efficiency and acceptance of the model-driven OO PLC programming process. The meta-model for the modAT4rMS including the structure implementation, behavior implementation and the connecting execution structure is shown in Fig. 1. For the corresponding specification and implementation of the UML class and state diagram and the basic activity diagram in an IEC 61131-3 environment, cf. [31], [32]. Also a comprised synthetic description of plcML from [32], on which part of modAT4rMS is based on, is given in the appendix. Subsequently the notation's elements and features are described using exemplary structure and behavior models of a stamping unit, created in a prototypical editor.

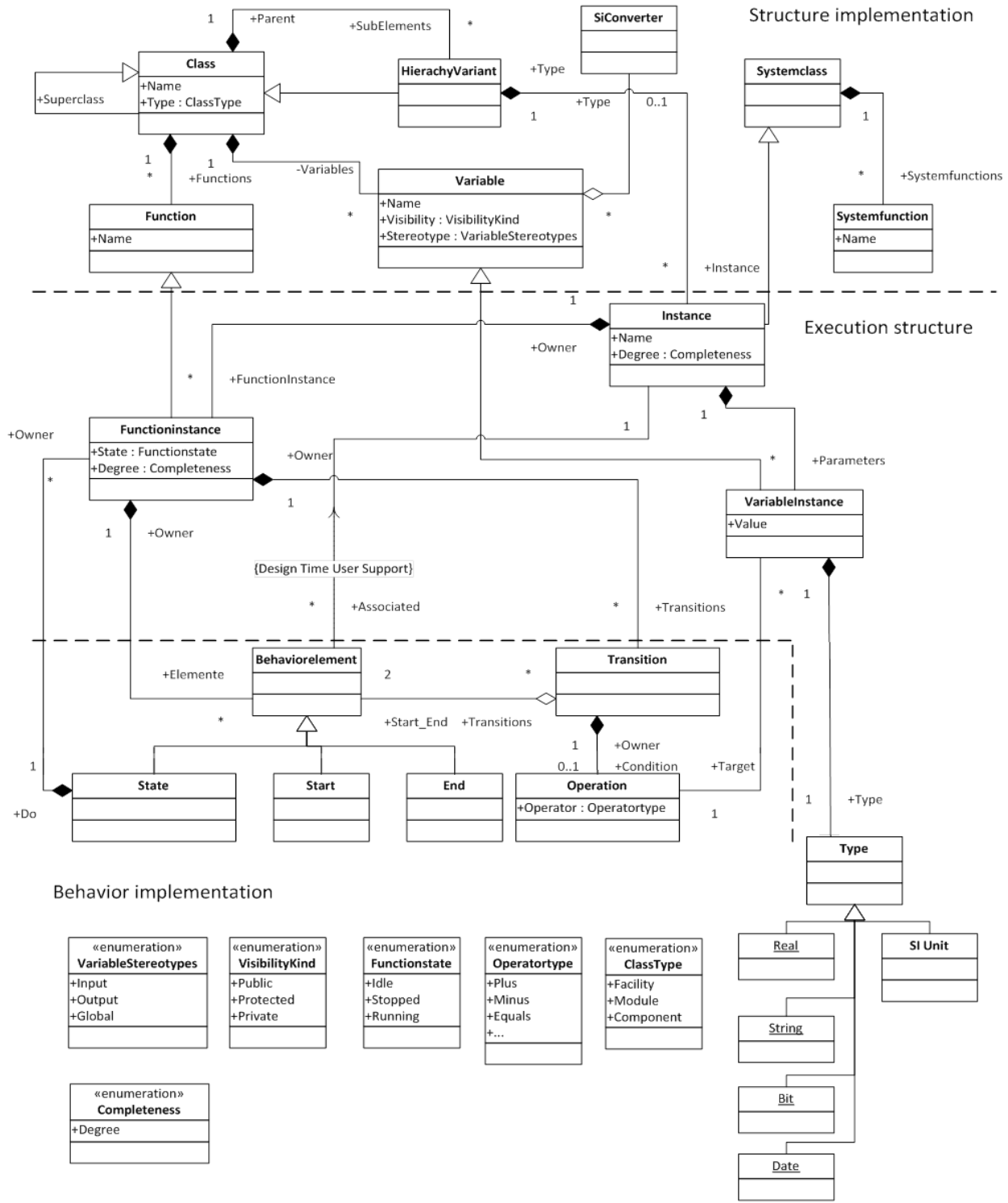

Fig. 1. Meta-model of modAT4rMS, upper third structure implementation, middle execution structure, lower third behavior implementation

*A. Structure model notation*

In the modAT4rMS structure model the UML class element is called block, adapted from the SysML, and is defined as shown in Fig. 2. It shows the extended block properties (additional variant and object name) as well as its integrated blocks, while in the upper third of Fig. 1 the underlying meta-model is shown. These properties were added to improve the usability of the structure model notation, by lowering the level of abstraction in comparison using UML standard diagrams, which would require at least the class and object diagram to provide the same information. This meta-model allows creating an inherited block in order to create variants of blocks. The inheritance corresponds to an inherited class of the class diagram or block of the SysML Block Definition Diagram and uses the meta-model's *HierarchyVariant* in order to provide information on its base block as well as the possibility to realize variants of blocks, cf. Fig. 1.

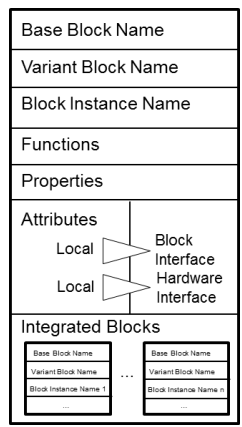

Fig. 2. modAT4rMS Block Definition

As a restriction to prevent the misuse of hardware descriptions through inheritance as software specifications and vice versa as well as in order to facilitate structure model comprehension, as deduced from the results in II.B, it is recommended to limit the inheritance depth to one. Through this limitation there is always only one base for different variants of a block, which can be easily recognized by the base block name. All blocks are specified in this manner, although the minimum functional level for a block is considered to be the sensor and actuator or component level. The level of the block is defined in the meta-model of modAT4rMS as the attribute *Type: Classtype* of the *Class*, cf. Fig. 1. A component block in the context of this paper represents the smallest unit integrated in a MS, e.g. sensors which supply only a signal value (digital/analog) and actuators which are being controlled by a single signal such as a simple valve. They are typically physically connected by a controller input or output. Aside from component level blocks, all other blocks can contain various other integrated blocks in their respective variant.

All blocks offer a functional interface through their methods and optionally their properties. This interface can only be accessed on the same hierarchy level or from the adjacent higher hierarchy level. Due to the strong hierarchy and the already large number of possible variants, the inheritance depth of one is deemed sufficient. Independent System functions are encapsulated in a *Systemclass*, cf. Fig. 1. As a limitation, the recommendation on inheritance depth is probably less capable to improve structure model comprehension for large scale models with many different variants of modules and little reuse, due to the possibly resulting high number of overall classes in such a scenario.

The *Systemclass* and its functions are exclusively accessible from every hierarchy level and therefore should be used only for block independent functions. Areas, where deeper inheritance levels would be required can be intercepted by a new conception for the base block or by an independent base in the form of an independent implementation specific inserted block and are thus left to the user. In order to get an impression of the modAt4rMS's structural meta-model and its connection to the corresponding behavior meta-model, cf. Fig. 1.

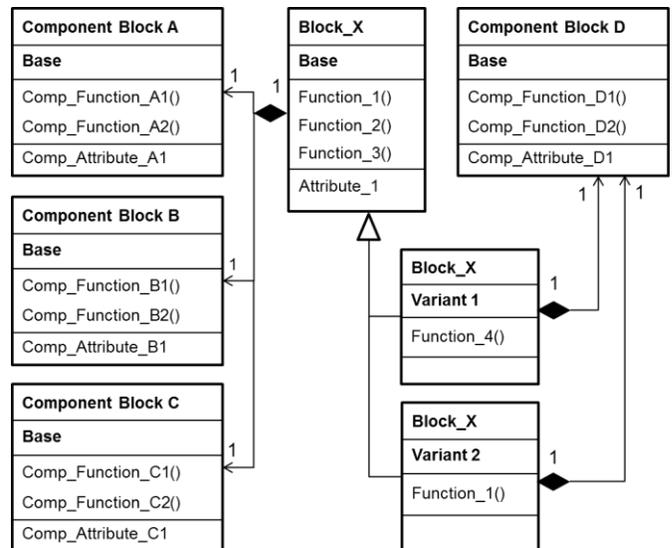

Fig. 3. Exemplary underlying modAT4rMS structure model for Block variant handling (block instance names and further irrelevant details not depicted)

An exemplary underlying structure model for a generic *Block_X* and its variants is shown in Fig. 3 and an approach for its creation is show in the following. In a best practice manner for modAt4rMS model creation, it is recommended to model the most common variant as the base block and to create all other variants through the addition or adaption of functions, attributes, and integrated blocks, cf. Fig. 3 *Block_X Variant1* and *Variant2*.

If the base block is chosen poorly at first and a new base block is in order, the modAT4rMS approach proposes changing the original base block into a variant block of this new base block. This is shown in Fig. 4 for the former base block *Block_X Variant3* and the new base for *Block_X*. All already existing variants of the original base block can become variants of the new base block, but keep their distinct functionalities, attributes and integrated blocks from before, cf. Fig. 4. Through this proposed variant handling approach the user is easily capable of keeping the inheritance structure clear and without the need for further structure modeling notations as the class diagram. The actual structure model created by the user is oriented on the UML object diagram as it is created as a specific instance for a specific machine. Its meta-model including the connection to the behavior model is defined as shown in Fig. 1.

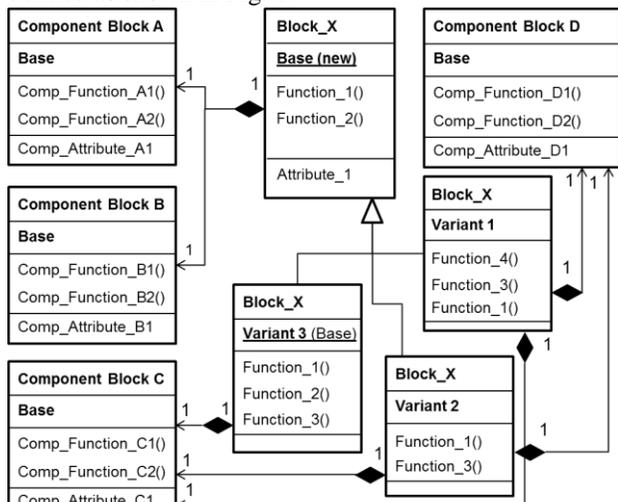

Fig. 4. modAT4rMS Block variant handling with base block change, original variant names in brackets, cf. Fig. 2 (irrelevant details not depicted)

### B. Behavior model notation

The behavior model is based on the fundamental ideas of the state and activity diagram of UML and respectively their plcML adaptions cf. [31], [32]. The modAT4rMS behavior notation describes sequences within the set of functions defined in the respective structure model. The connection between the structure and behavior model are defined via the execution structure depicted in Fig. 1. The graphical representation of the modAT4rMS adapts parts of the UML sequence diagram, as it uses object specific columns divided by swim lanes, cf. Fig. 5. In contrast to the sequence diagram, the modAt4rMS behavior notation is capable of describing all scenarios of the modelled function. In order to create a clear connection between the structure and behavior model, the available objects are based on the structure model of the function's host block. Additionally, the columns are structured into separate fields. In general, each field of the behavior diagram can be used to call one function (symbolized with an oval frame, cf. Fig. 5) or to access an attribute (symbolized with a rectangular frame, cf. Fig. 5) of the corresponding block, as defined in their structure model. The underlain meta-model defines states corresponding to the definition of the *Behaviorelement*, cf. Fig. 1, respectively the state element of the plcML, cf. [31], [32]. The definitions in terms of cyclic execution, as well as priorities and execution routines such as "entry", "Do", "Exit" are also taken completely over from plcML, cf. [31], [32].

Functions and operations can also access the attributes of blocks (symbolized by a surrounding additional frame around the function call or operation and the corresponding attributes/parameters, as labeled in Fig. 5). Operations can use this feature only with for condition forming, cf. Fig. 5 upper additional frame, whereas functions can use it if their structure model defines one or more function call parameters. A function call creates a *Functioninstance* and can contain several states in which operations or other functions are called. An operation is owned by an transition and either manipulates a *Variableinstance*, e.g. Signal:=1 or describes a condition,

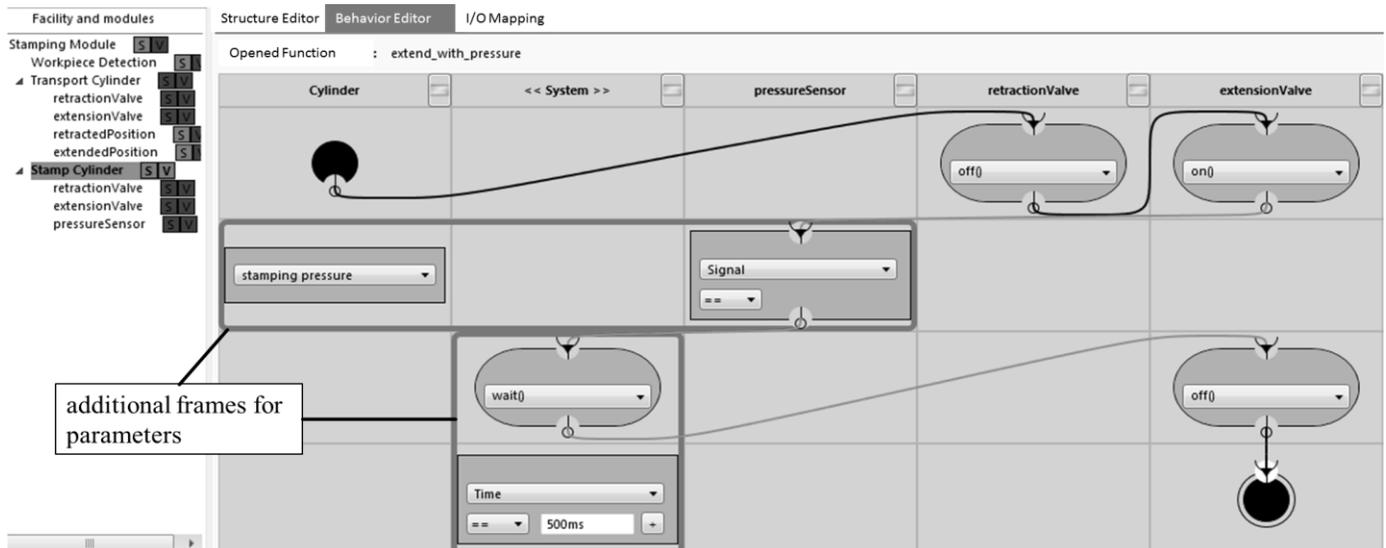

Fig. 5. Exemplary behavior model for a stamping cylinder function in the developed editor

e.g. pressureSensor.Signal==1, cf. Fig. 5.

Start and end points are reused from the plcML activity diagram or respectively the plcML state chart diagram. In regard to the cyclic nature of PLC program processing, all nodes in the behavior model are multicycle nodes, cf. [31], [32]. In short, a new plc cycle is started after the operation/function is finished in the active node before either the operation/function is repeated in the next cycle or an enabled outgoing transition leads to a new active node.

Two types of transitions have been defined in modAT4rMS in order to connect start/end points: function calls and attribute operations, whereas the transition meta-model is reused from the plcML, cf. [31], [32]. The presentation of the necessary condition is achieved by a separate element. The default transition type for function calls is the Completion Transition, adapted from plcML, which executes the following function call only after the previous function has ended. The other type of transition uses attributes as switching conditions, e.g. sensor signals. For this condition an attribute can either be compared to a fixed value, a function call parameter (cf. Fig. 5) or another attribute at the same hierarchy level. The use of the second type of transition makes it necessary to define how the conditions preceding function call is handled. It is possible to interrupt the function and continue with the next call at this point; another possibility is to interrupt the function and start the function from its beginning when it is next called or to let the function continue.

*C. modAT4rMS modeling example*

In order to evaluate the developed modAT4rMS approach, a prototypical model editor was created, which was used to create the modeling example that is described in the following. It applies the modAT4rMS approach for an automation example of a stamping unit. Fig. 6 shows the underlying structure model of the example, whereas Fig.7 shows the modAt4rMS depiction of the Stamp module and a part of the developed graphical user interface, used in the evaluation, cf. IV.

In order to facilitate the creation of the users' mental models of the system and therefore the creation of the software model, technical sketches or pictures of the corresponding hardware modules can optionally be put in the editor's background and only integrated blocks of the next lower hierarchy level of the block are shown. The abstraction level of the structure diagram of modAT4rMS is reduced by giving the block name, variant name and instance name in one diagram additionally to its attributes, interfaces and functions. This aims to lower the subjective workload and to rise the efficiency of the modelling process.

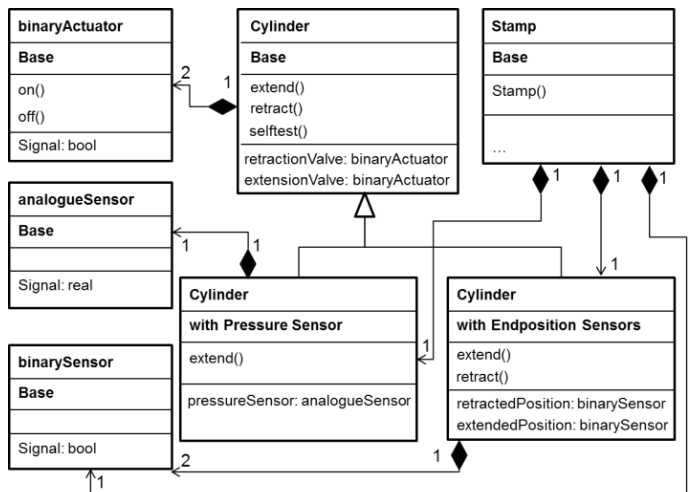

Fig. 6. Simplified underlying structure model of the modAT4rMS modeling example (not visible in the modAT4rMS editor)

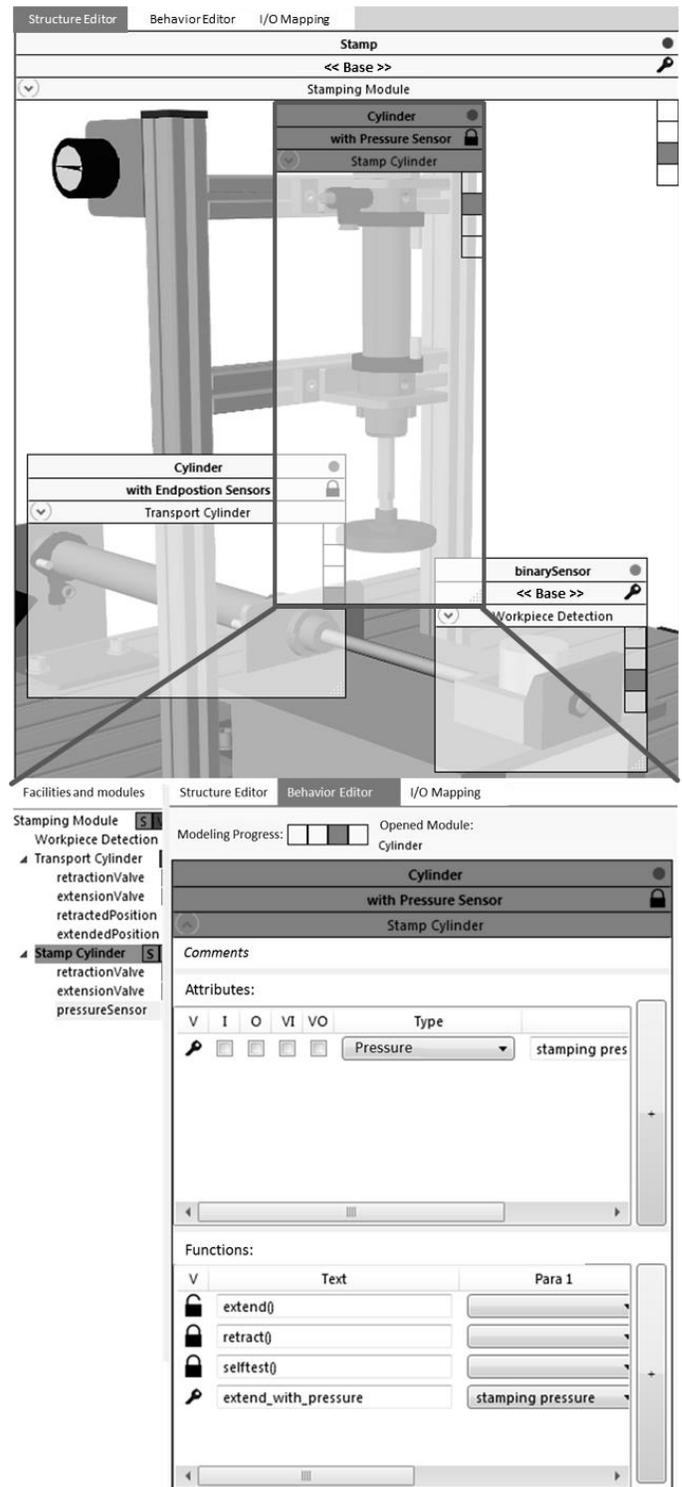

Fig. 7. Exemplary structure model for the *Stamp* unit in the developed editor

For the *Stamp* block, shown in Fig., this means two different pneumatic cylinder variants created from the same base cylinder are integrated as well as one *binarySensor* block, cf. also Fig. 7.

The *Transport Cylinder* is a cylinder variant, which possesses two additional binary end position sensors, cf. Fig. 6. The *Stamp Cylinder* variant includes an analogue pressure sensor, cf. Fig. 7 lower half and Fig. 6 *pressureSensor*. The base cylinder block (cf. Fig. 6, not shown in Fig.7) integrates two *binaryActuator* blocks, which are controlled by the cylinder functions *extend()*, *retract()*. Both cylinder block variants inherit the two valves from the base cylinder, cf. Fig 6. The functional interface in the form of a collection of functions (*extend()*, *retract()*, *selftest()*) also is inherited, which is symbolized through a padlock symbol in front of the function names, cf. Fig.7 lower half. If a function of the block variant has been altered, the padlock is opened, if not it is closed. Additionally to his inherited functions the variant includes a new function, which can be recognized by a key symbol in front of the function name. In combination with the limitation of inheritance depth to one, this enables a quick overview of structure dependence between base and variant blocks without the need to use an additional structure notation. Continuing the modeling example of the stamping unit, the behavior model for the adapted *extend()* function of the stamping cylinder will be discussed following in the and is shown in Fig. 5.

The functions and attributes of these blocks can be used to create the behavior model for functions of the *Stamp Cylinder*, like the *extend()* function, cf. Fig. 5. All integrated blocks are shown in the top row of the behavior model, divided by swim lanes. This facilitates the representation and respectively recognition of the structural context of the corresponding behavior model, as demanded in the results of II.B. Additional fields, always present in the top row, are the block itself and so-called system or library functions. The system or library functions are exempt from all inheritance and access limitations defined in this paper. This is necessary to ease the use of programming environment tools such as logic operations, timers and independently implemented standalone software functions. On the one hand these cannot be assigned to any block and on the other hand, it makes no sense to implement them anew in the modAT4rMS diagrams, rather than reusing their high level language implementations. Therefore, these functions are handled in a black-box manner, only offering an interface for the user with the possibility to transfer parameters for the function call.

IV. EVALUATION

In order to evaluate the new modAT4rMS notations an experimental study was conducted, comparing it to the state of the art IEC 61131-3 FBD and the domain specific plcML notation. The study took place from November 2012 until January 2013 at the Technische Universität München (one day training, one day experimentation per group). The sample of 168 participants consisted of five school classes (three beginner 2nd year classes, two intermediate 3rd year classes) from a vocational school for production engineering in Munich with specialization in mechatronics, as these apprentices form the relevant group of novice and intermediate PLC programmers in Germany, and two expert classes consisting only of industrial practitioners with several years of work experience from a school for state certified technicians in Munich. Programming performance was measured by the level of task completion, cf. [4]. In addition user acceptance aspects, the subjective ease of use, frustration and usability were measured using scaled questions on the aspects defined in [33], [36]. Similar to the previous studies in II.B, the task completion measure was used to evaluate the programming performance respectively the effectiveness. The analysis of the model and code was carried out by two evaluators per notation in order to verify the objectivity of the evaluation method through the level of agreement using Cohen's Kappa-coefficient [38]. This provides an objective basis as a prerequisite for a correct evaluation. In order to analyze the results, the inferential statistics method of multivariate analysis of variance (MANOVA) was used in combination with post-hoc Scheffé tests, cf. [39]. In principle, this test allows analyzing the variance of results between more than two groups of results in order to test for significant differences between the different group-pairs directly, e.g. between plcML and FBD, modAT4rMS and FBD, etc. For the probability of results occurring by chance (p-value), the threshold of 5% for significant and <1% for highly significant differences were used. The automation problem for testing the approach was adapted from industrial practice, where it is used as a final exam for PLC programmers after at least three years of training. In cooperation with PLC programmers from industrial contacts (e.g. Liebherr-Verzahntechnik GmbH) and PLC and automation teachers of the vocational school of the participants the test was validated, to be generalizable for automation problems in industrial practice and in teaching. The given task includes the creation of a control program model for a MS consisting of five transport portals and four processing units (overall using 100 I/Os). This task was divided in 3 subtasks: 1a (creation of portals 1-4 and processing units 1-3, 74 I/Os), 1b (creation of portal 5 and processing unit 4 as specialized variants, 26 I/Os) and 1c (modification of functions calls with parameters for portals 1-4 and processing units 1-4) for which 60 minutes (1a), 50 minutes (1b) and 25 minutes (1c) were given.

Further detail on the experimental design and a full description of all further measured variables and results will be published in papers focusing on these additional aspects as soon as all necessary data is processed.

*A. Results on programming performance*

All results on programming performance (task completion) for each test task are presented in relation to the experience level of the participants, cf. Fig. 6, Fig. 7, Fig. 8. The results on the overall programming performance in the model creation task 1a showed highly significant effects ($p < 0.001$) in favor of modAT4rMS, cf. Fig. 6.

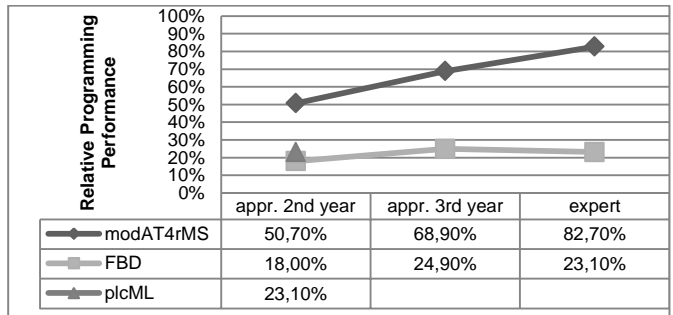

Fig. 6. Results on relative overall programming performance in the model creation subtask 1a

There was a significant interaction effect between notation and experience. As depicted in Fig. 6, the more experienced groups profited more of the modAT4rMS notations than the lesser experienced; although no prior experience in modAT4rMS modeling was present. The results for the overall programming performance in the model creation and reuse task 1b showed also highly significant effects (p < 0.001) in favor of modAT4rMS, cf. Fig. 7.

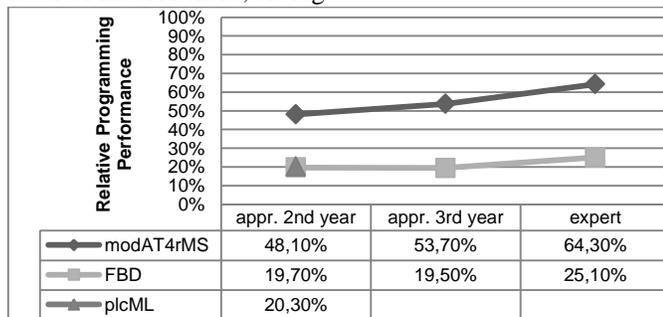

Fig. 7. Results on relative overall programming performance in the model creation and reuse subtask 1b

Between the IEC 61131-3 FBD and the plcML no significant difference occurred and no significant interaction effects between experience and programming performance were measured.

Fig. 8 shows the results of subtask 1c, which required modifying existing functions to become parameterized functions and to integrate them into the existing control programs. The MANOVA showed no significant differences between the notations alone with regard to programming performance for this subtask (p = 0.549), but a significant interaction effect between the notation and experience levels (p=0.032), in favor of the expert level using FBD and the 3rd. year level using modAt4rMS, cf. Fig. 10.

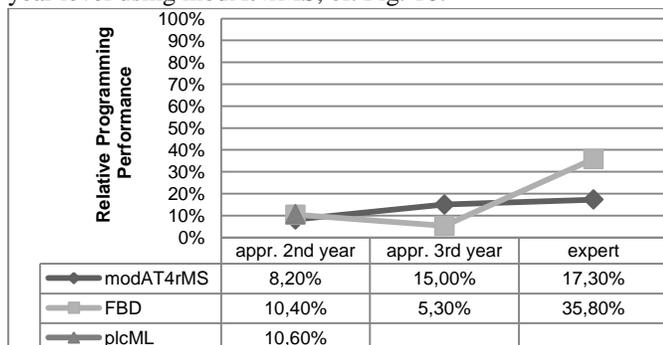

Fig. 8. Results on relative overall programming performance in the modifying existing functions subtask 1c

## B. Results on the subjective usability: ease of use, frustration, workload

The workload differed significantly (p < 0.001) with respect to the used programming notation. The post-hoc Scheffé test showed that the groups who worked with the notations of modAT4rMS experienced highly significantly less stress than the comparison notations of the IEC 61131-3 FBD (p <0.001) and the plcML (p = 0.001). No differences were found between IEC 61131-3 and the plcML (p = 0.999). Furthermore, the experience of the subjects showed no significant effect on workload (p = 0.469). Also, no interaction effects (p = 0.837) were observed.

Just as with workload, the analysis of the frustration data showed highly significant results (p < 0.001). The evaluation of the post-hoc Scheffé tests showed again that subjects who worked with the notations of modAT4rMS were highly significantly less frustrated than the subjects with the comparison notations of IEC61131-3 FBD (p < 0.001) and the plcML (p = 0.001). No significant differences were found between IEC61131-3 FBD and the plcML (p = 0.433). Furthermore, the experience of the subjects showed no significant effect on their stated frustration (p = 0.714). Also, no significant interaction effects (p = 0.134) were observed.

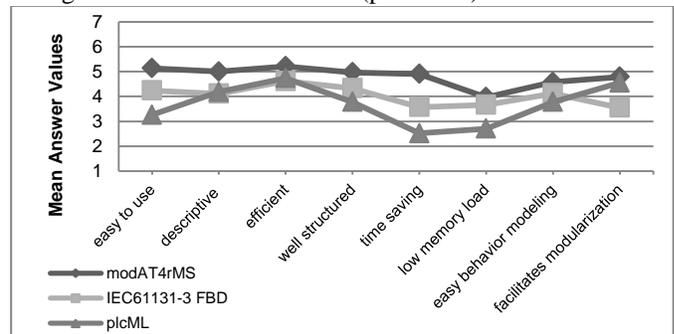

Fig. 9. Results on subjective feedback questionnaire (scale from 1 = statement not fitting at all to 7 = statement completely fitting)

The mean subjective questionnaire values for the used modeling notations are shown in Fig. 9. The statements were in the form of "The used notation is/causes …". Overall modAT4rMS received the most positive feedback, whereas plcML and IEC 61131-3 FBD received mixed feedback, cf. Fig. 9.

## C. Discussion of the results

The programming performance of subjects when creating a new PLC program model indicate clear advantages of the modAT4rMS notations, cf. Fig. 6, Fig. 7. Because of the interaction effects (experience * notation), the better programming performance of the subjects in programming the system behavior cannot be considered separately. However, the results show clear performance differences in favor of modAT4rMS notations. Furthermore, the results on the programming performance in terms of the experience in PLC programming show mostly a particularly positive effect on the modAT4rMS notations, cf. Fig. 6. Only for the subtask on modifying existing functions 1c, FBD seems to profit stronger than modAt4rMS from the experts' experience in IEC 61131-3.The subjective answers on workload, frustration and further usability aspects all show a highly significant positive effect of the modAT4rMS notations. Overall the gained results show that the developed notations offer an objective and subjective significantly better usable alternative to existing notations in order to promote model-driven OO programming for PLCs in the MS field. The results on the compared plcML and modAt4rMS notations, both with the same formal basis, also show that an improvement in usability does not necessarily require a different formal basis.

## V. Conclusion and Outlook

In this paper the modAT4rMS approach for the creation of well-structured and comprehensible PLC software was presented. The goal of this new approach was to facilitate the use of OO programming in machine and plant automation by providing domain specific modeling notations. Problems during the structure and behavior creation were identified in previous studies and were taken into account in the development of the modAT4rMS. The final modAT4rMS notations significantly eased former problems as shown in the results of the empirical evaluation. In all but one of the presented programming subtasks the modAT4rMS proved significantly superior and the subjective results are in favor of the developed approach as well.

Based on these results, the proposed block and component-based structure of the control software should also provide a good basis to communicate with the domain of mechanical and process engineering as well with PLC programmers. As the data analysis of the conducted evaluation experiment is still ongoing, further results on the proposed structure abstraction layer, occurred errors etc. will be presented in future work. As the modAT4rMS meta-model is based on the plcML, its notations can be transformed to real time capable IEC 61131-3 code and could possibly be integrated into an IEC 61131-3 programming environment in the future.

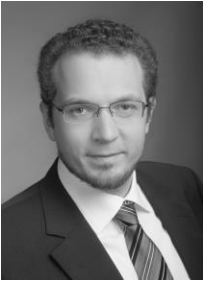

**Martin Obermeier** received the Dipl.-Ing. degree in mechanical engineering (focus on Control Theory and Information Technology) from Technische Universität München, Germany, in 2009. He is currently working towards the Ph.D. degree at the Institute of Automation and Information Systems, Technische Universität München, Germany.

Mr. Obermeier does research on the design and evaluation of model based programming software for automation systems in regard to model quality, modularity aspects as well as usability.

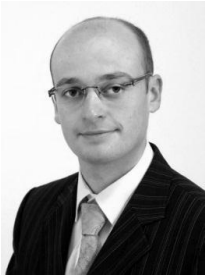

**Steven Braun** graduated in mechanical engineering (focus on Mechatronics and Information Technology) from Technische Universität München, Germany, in 2008 and received the PhD. degree in mechanical engineering in 2014. He was working towards the Ph.D. degree at the Institute of Automation and Information Systems, Technische Universität München, Germany.

Mr. Braun does research in the design and evaluation of model based programming software for automation systems.

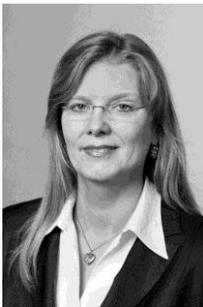

**Birgit Vogel-Heuser** (M'04, SM'12) graduated in electrical engineering and received the PhD. degree in mechanical engineering from the RWTH Aachen in 1991.

She worked nearly ten years in industrial automation for machine and plant manufacturing industry. After holding different chairs of automation in Hagen, Wuppertal and Kassel she is since 2009 head of the automation and information systems institute at the Technische Universität München. Her research work is focused on modeling and education in automation engineering for hybrid process and heterogeneous distributed and intelligent systems using a human centered approach.

Prof. Vogel-Heuser is member of the GMA (NMO IFAC). She received four awards, i.e. Special Award of the Initiative D21 Women in Research (2005), Borchers Medal of the RWTH Aachen (1991), GfR Sponsorship Award (1990) and Adam Opel Award (1989).

APPENDIX. META-MODEL CONNECTION BETWEEN PLCML AND MODAT4RMS

In the following a comprised view on the plcML-Statecharts are given and a synthetic description on the connection between the plcML and modAt4rMS is given.

plcML-Statecharts are defined in terms of a UML profile. Therefore, they support a large set of modeling elements specified in the UML standard. Figure 12 depicts the resulting UML meta model (shaded meta-classes are defined by the profile, non-shaded by UML).

The connection between the modAT4rMS meta-model and the plcML and UML is comprised in Fig. 13 and Fig. 14. The UML elements used in plcML are shown in table 1. In Fig. 13 a comprised view on the parts of the modAt4rMS structure notation adapted from plcML and UML is given. As shown *Class*, *Function, VariableStereotypes* and *VariableInstance* are adapted from the plcML meta-model. In Fig. 14 a comprised view on the parts of the modAt4rMS behavior notation adapted from plcML and UML is given. As shown, *State*, *Transition, Operation, Start* and *End* are adapted from the plcML, respectively from the UML meta-model. All parts, new and adapted, of the modAt4rMS meta-model and their relationships are depicted in Fig. 1.

TABLE 1: UML ELEMENTS USED IN PLCML, CF. [32]

| plcML element | UML element | UML package |
|---|---|---|
| expansion mechanisms | profiles | - |
| data types | auxiliary constructs | PrimitiveTypes |
| class diagram | classes | Kernel |
| | | Dependencies |
| | | Interfaces |
| state chart diagram | common behaviors | BasicBehaviors |
| | | SimpleTime |
| | | Communications |
| | StateMachines | BehaviorStateMachines |
| | Activities | FundamentalActivities |

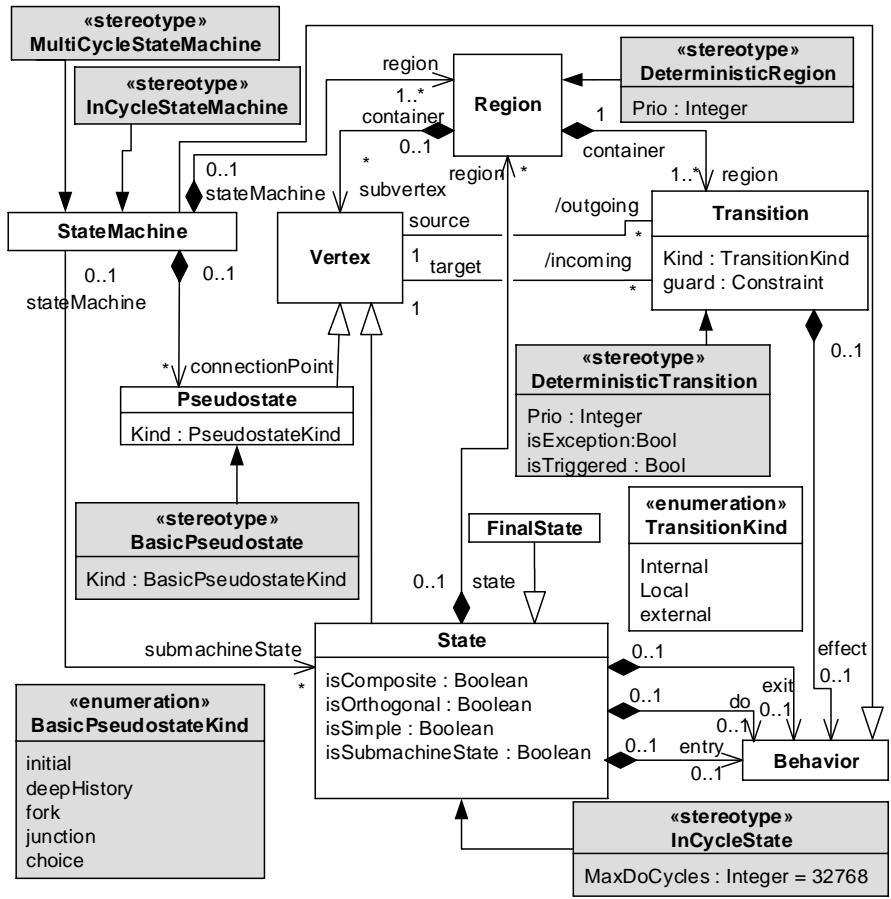

Fig. 10. Meta-model for plcML-Statecharts (shaded meta-classes are defined by the plcML profile, non-shaded by UML)., cf. [31]

Fig. 11. meta-model elements of the modAT4rMS structure notation adapted from UML and plcML

Fig. 12. Meta-model elements for the modAT4rMS behavior notation adapted from UML and plcML